\documentclass[10pt]{article}
\usepackage{graphicx,psfrag,color,amsfonts,amsmath,upgreek}
\usepackage{stmaryrd,cancel,mathrsfs,amssymb}
\DeclareMathAlphabet{\mathpzc}{OT1}{pzc}{m}{it}
\usepackage[colorlinks=true,linkcolor=blue, citecolor=blue]{hyperref}
\usepackage[firstinits=true,doi=true,isbn=false,url=false,sorting=nyt,eprint=false]{biblatex}

\begin{document}
\title{Energy expense via lattice wave emission for mode III brittle fracture in square, triangular, and hexagonal lattices}
\author{Basant Lal Sharma\thanks{Department of Mechanical Engineering, Indian Institute of Technology Kanpur, Kanpur, U. P. 208016, India ({\tt bls@iitk.ac.in}).}}

\maketitle

\begin{abstract}
The mode III fracture problem for a hexagonal lattice is discussed and compared with square and triangular lattices. 
\end{abstract}

\section*{Introduction}
\begin{figure}[h]
\begin{center}
{\includegraphics[width=.7\linewidth]{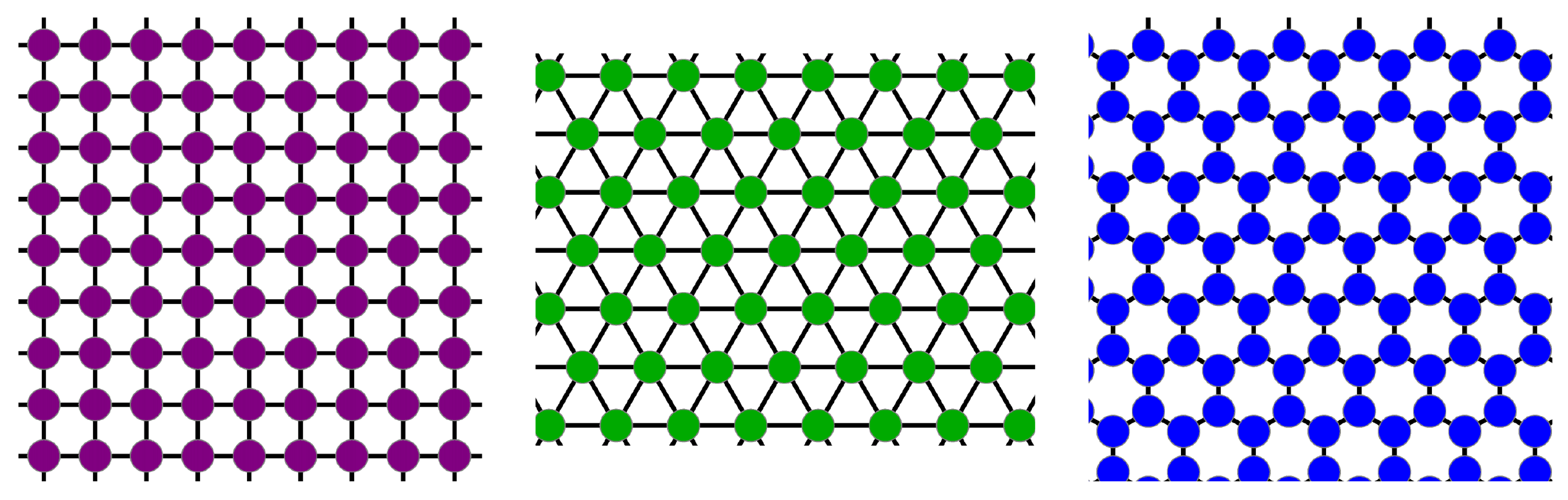}}
\caption{Square, triangular, and hexagonal lattices with nearest neighbour bonds.}
\label{Fig1}
\end{center}
\end{figure}

A well known discrete mechanical model of crack \cite{Slepyan1981a,Slepyanbook} (see also \cite{Marder, marder2}) has been adapted in problem formulation for mode III brittle fracture in a hexagonal lattice. See Fig. \ref{Fig1} for an illustration of the two dimensional lattice models including square and triangular.

It is assumed that each particle, in the two dimensional lattice is connected with its three nearest neighbors by linearly elastic identical (massless) bonds. The bonds have a spring constant ${K}$ and an equilibrium length ${b}$. 
The results of \cite{Slepyan1981a, Slepyanbook} for square lattice and those of \cite{Marder} for triangular lattice, are briefly recollected in the appendices \ref{appSaniso} and \ref{appTaniso}, respectively. In addition to the well studied models, an anisotropy parameter is also incorporated.

\subsection*{Notation}
Let $\mathbb{Z}$ denote the integers.
Let $\mathbb{Z}^2$ denote the grid $\mathbb{Z}\times\mathbb{Z}$.
Let $\mathbb{R}$ denote the real numbers and $\mathbb{C}$ the complex numbers.
In this manuscript, $\upxi$ typically denotes the complex variable of Fourier transform. 
The letter ${{{\mathscr{H}}}}$ stands for the Heaviside function: ${{{\mathscr{H}}}}(m)=0, m<0$ and ${{{\mathscr{H}}}}(m)=1, m\ge0$.
The square root function, $\sqrt{\cdot}$, has the usual branch cut in the complex plane running from $-\infty$ to $0$.
${{\updelta}}_{p, q}$ equal to $1$ if $p=q$ and zero otherwise is Kronecker delta function. 
Other physical entities are defined in the analysis according to their relevance.

\section{Mode III crack in square lattice model}
\label{appSaniso}
Suppose that a mode III crack is moving at a constant velocity with magnitude ${{\mathtt{V}}}={{\mathcal{V}}}{c_s}>0, $ in the crack plane ${\mathcal{P}}$ between ${\mathtt{y}}=-$ and ${\mathtt{y}}=-1.$ 
The crack (scaled) velocity ${\mathcal{V}}=1$ corresponds to the physical velocity ${\mathtt{V}}={c_s}$. A piecewise linear inter particle force law 
between the nearest neighbours is assumed. 
The equation of motion for the particle located at $({\mathtt{x}}, {\mathtt{y}})$, based on the classical (Newtonian) mechanics, is directly affected by the presence of crack whenever ${\mathtt{y}}=-{{\frac{1}{2}}}\pm{{\frac{1}{2}}}$. The lattice fracture criterion is \cite{Marder, Slepyan1981a} 
\begin{align}
\llbracket{{\mathtt{u}}}\rrbracket^{{\mathcal{P}}{\mathrel{\text{\ooalign{$\downarrow$\cr$\uparrow$}}}}}({\mathtt{x}}, {{\mathtt{t}}})={{\mathtt{v}}_{{\mathrm{c}}}} \text{ at }{{\mathtt{t}}}=\frac{\mathtt{x}}{{\mathcal{V}}}, {{\mathtt{x}}}\in{\mathbb{Z}}.
\label{sq_fracbond}
\end{align}
with $\llbracket {{\mathtt{u}}}\rrbracket^{{\mathcal{P}}{\mathrel{\text{\ooalign{$\downarrow$\cr$\uparrow$}}}}}({\mathtt{x}}, {{\mathtt{t}}})={{\mathtt{u}}}_{{\mathtt{x}}, 0}-{{\mathtt{u}}}_{{\mathtt{x}}, -1}$.
The total displacement satisfies the equation
\begin{align}
\frac{d^2}{d{\mathtt{t}}^2}{{\mathtt{u}}}_{{\mathtt{x}}, {\mathtt{y}}}&=\triangle{{\mathtt{u}}}_{{\mathtt{x}}, {\mathtt{y}}}+{{\mathscr{H}}}({{\mathcal{V}}}{{\mathtt{t}}}-{\mathtt{x}})({{\updelta}}_{{\mathtt{y}}, 0}-{{\updelta}}_{{\mathtt{y}}, -1})\llbracket {{\mathtt{u}}}\rrbracket^{{\mathcal{P}}{\mathrel{\text{\ooalign{$\downarrow$\cr$\uparrow$}}}}}({\mathtt{x}}, {{\mathtt{t}}}), ({\mathtt{x}}, {\mathtt{y}})\in{{\mathbb{Z}^2}}.
\label{sq_modeIIIcrackeq1}
\end{align}
A subcritical crack speed is assumed: $0 < {\mathtt{V}} < {{c_s}}.$ 

For the considered steady-state problem a moving coordinate, ${\mathtt{z}}=\mathtt{x} - {\mathcal{V}}{\mathtt{t}}$ is introduced. Assuming ${{\mathtt{u}}}_{\mathtt{x}, \mathtt{y}}(t)={{\mathtt{u}}}_{\mathtt{y}}({\mathtt{z}})$ equation \eqref{sq_modeIIIcrackeq1} can be rewritten in the form 
\begin{align}
{\mathcal{V}}^2\frac{d^2}{d{\mathtt{z}}^2}{{\mathtt{u}}}_{{\mathtt{y}}}({\mathtt{z}})={{\mathtt{u}}}_{{\mathtt{y}}}({\mathtt{z}}+1)+{{\mathtt{u}}}_{{\mathtt{y}}}({\mathtt{z}}-1)+{{\mathtt{u}}}_{{\mathtt{y}}+1}({\mathtt{z}})+{{\mathtt{u}}}_{{\mathtt{y}}-1}({\mathtt{z}})-4{{\mathtt{u}}}_{{\mathtt{y}}}({\mathtt{z}}), 
\label{sq_modeIIIcrackeq2}
\end{align}
away from the crack. The Fourier transform on ${\mathtt{z}}$ for ${\mathtt{y}} \ge 0$ leads to a general solution of the form
\begin{subequations}\begin{eqnarray}
{{\mathtt{u}}}^F_{\mathtt{y}}({\upxi})&=&{{\mathtt{u}}}^F({\upxi}){\lambda}^{\mathtt{y}}({\upxi}), \\
\text{with }{{\mathtt{u}}}({\mathtt{z}})&=&{{\mathtt{u}}}_0({\mathtt{z}}), {\lambda}({\upxi})=\frac{{\mathpzc{r}}({\upxi}) - {\mathpzc{h}}({\upxi})}{{\mathpzc{r}}({\upxi}) + {\mathpzc{h}}({\upxi})}~(|{\lambda}| \le 1),\label{sq_modeIIIcrackeq3}\\
{\mathpzc{h}}^2({\upxi})&=&2(1- \cos {\upxi})+(0+i{\upxi}{\mathcal{V}})^2, {\mathpzc{r}}^2={\mathpzc{h}}^2 +4.
\label{sq_modeIIIcrackeq4}
\end{eqnarray}\label{sq_uansatz}\end{subequations}
Accounting for the skew-symmetry, due to far-field stress ${\upsigma}$,
the Fourier transform of the equation at $\mathtt{y}=0$ yields
\begin{subequations}\begin{eqnarray}
{{\mathtt{u}}}_{+}({\upxi}) + {\mathpzc{L}}_{\mathfrak{S}}({\upxi}){{\mathtt{u}}}_{-}({\upxi})&={{\frac{1}{2}}}(1-{\mathpzc{L}}_{\mathfrak{S}}({\upxi})){{\mathfrak{q}}},\label{sq_modeIIIcrackeq8}\\
\text{where }
{{\mathfrak{q}}}=\frac{{\upsigma}}{i{\upxi}+{0+}}, {{\mathpzc{L}}}_{\mathfrak{S}}=\frac{{\mathpzc{h}}}{{\mathpzc{r}}}.\label{sq_Lz}
\end{eqnarray}\label{sq_WHL}\end{subequations}
Following \cite{Slepyan1981a}, a limit is considered
as $\epsilon\to0+, \frac{{\upsigma}}{i{\upxi}+{\epsilon}}=\frac{{K}\sqrt{2\epsilon}}{i{\upxi}+{\epsilon}}$, where ${K}$ is the stress intensity factor for the macroscopic stress field around the crack tip.

A slight generalization involves the horizontal bonds of spring constant ${\upchi}$. The right side of \eqref{sq_modeIIIcrackeq2} becomes ${\upchi}{{\mathtt{u}}}_{\mathtt{y}}({\mathtt{z}}+1)+{\upchi}{{\mathtt{u}}}_{\mathtt{y}}({\mathtt{z}}-1)+{{\mathtt{u}}}_{\mathtt{y}+1}({\mathtt{z}})+{{\mathtt{u}}}_{\mathtt{y}-1}({\mathtt{z}})-(2+2{\upchi}){{\mathtt{u}}}_{\mathtt{y}}({\mathtt{z}})$ and \eqref{sq_modeIIIcrackeq4} becomes
\begin{align}
{{\mathpzc{h}^2}}({\upxi})=2{\upchi}(1- \cos {\upxi})+(0+i{\upxi}{\mathcal{V}})^2.
\label{sq_h2chi}
\end{align}
The crack (scaled) velocity ${\mathcal{V}}=\sqrt{{{\upchi}}}$ corresponds to the physical velocity equal to the speed of `sound', i.e. ${\mathtt{V}}={c_s}$.
Let
\begin{align}
\mathcal{C}_{\mathfrak{S}}=\sqrt{{{\upchi}}}.
\label{sqsound}
\end{align}

It can be shown that \eqref{sq_WHL} involves an 
index ${{\mathpzc{L}}}_{\mathfrak{S}}=0$, ${{\mathpzc{L}}}_{\mathfrak{S}}(\pm\infty)=1,$
and
$${{\mathpzc{L}}}_{\mathfrak{S}}({\upxi})\sim \frac{1}{2}\sqrt{\mathcal{C}_{\mathfrak{S}}^2-{\mathcal{V}}^2}\sqrt{0+i{\upxi}}\sqrt{0-i{\upxi}}$$ as ${\upxi}\to0$.
The decomposition of kernel \eqref{sq_WHL} as
\begin{align}
{{\mathpzc{L}}}_{\mathfrak{S}}={{\mathpzc{L}}}_{{\mathfrak{S}}+}{{\mathpzc{L}}}_{{\mathfrak{S}}-},
\end{align}
where
\begin{align}
{{\mathpzc{L}}_{\mathfrak{S}}}_\pm({\upxi})&=\exp(\mp\frac{1}{2\pi i} \int_{\mathbb{R}}\frac{\log {{\mathpzc{L}}_{\mathfrak{S}}}({s})}{{\upxi}-{s}}d{s}), {\upxi}\in{\mathbb{C}}, \Im{\upxi}\gtrless\mp0,\label{sq_LKpmexp}
\end{align}
holds that gives the following asymptotic expressions (${\mathcal{V}}<\mathcal{C}_{\mathfrak{S}}$): 
\begin{subequations}
\begin{align}
{{\mathpzc{L}}}_{{\mathfrak{S}}+}({\upxi}) &\to 1~({\upxi} \to+i\infty),\quad {{\mathpzc{L}}}_{{\mathfrak{S}}-}({\upxi}) \to 1~({\upxi} \to-i\infty), \\
{{\mathpzc{L}}}_{{\mathfrak{S}}+}({\upxi}) &\to (\frac{1}{2}\sqrt{\mathcal{C}_{\mathfrak{S}}^2-{\mathcal{V}}^2})^{1/2}(0-i\upxi)^{1/2}{{\mathfrak{C}}}_{\mathfrak{S}}~({\upxi} \to 0),\\{{\mathpzc{L}}}_{{\mathfrak{S}}-}({\upxi}) &\to (\frac{1}{2}\sqrt{\mathcal{C}_{\mathfrak{S}}^2-{\mathcal{V}}^2})^{1/2}(0+i\upxi)^{1/2}{{\mathfrak{C}}}^{-1}_{\mathfrak{S}}~({\upxi} \to 0), 
\label{sq_Lzplim}
\end{align}
\end{subequations}
where 
\begin{align}
{{\mathfrak{C}}}_{\mathfrak{S}}&=\exp(\frac{1}{\pi}\int_0^{+\infty}\frac{\arg{{{\mathpzc{L}}}_{\mathfrak{S}}({\upxi})}}{{\upxi}}d{\upxi}).
\label{sq_ConL}
\end{align}
Reverting back to the case ${\upchi}=1$, further proceeding by the application of the Wiener--Hopf method \cite{Slepyanbook}, it is found that
\begin{subequations}
\begin{align}{\mathpzc{L}}_{{\mathfrak{S}}+}^{-1}{{\mathtt{u}}}_{+} + {\mathpzc{L}}_{{\mathfrak{S}}-}({\upxi}){{\mathtt{u}}}_{-}
&=\pi \frac{{K}}{\sqrt{i}(1-{\mathcal{V}}^2)^{1/4}}\delta({\upxi}).\end{align}Finally, the solution is
\begin{align}
{{\mathtt{u}}}_{+}({\upxi})={{\mathpzc{L}}}_{{\mathfrak{S}}+}({\upxi})\frac{\sqrt{i}}{2} \frac{{K}}{(1-{\mathcal{V}}^2)^{1/4}}\frac{1}{0i+{\upxi}}\\
{{\mathtt{u}}}_{-}({\upxi})=-{{\mathpzc{L}}}_{{\mathfrak{S}}-}^{-1}({\upxi})\frac{\sqrt{i}}{2} \frac{{K}}{(1-{\mathcal{V}}^2)^{1/4}}\frac{1}{-0i+{\upxi}}.
\end{align}
\end{subequations}
By the fracture condition,
\begin{align}
\lim_{s\to\infty}u_{+}(i{s})s=\frac{1}{2}\frac{{K}}{(1-{\mathcal{V}}^2)^{1/4}{{\mathfrak{C}}}_{\mathfrak{S}}}=\frac{1}{2}\sigma_c.
\end{align}
The energy required to snap the bond each time the crack advances by unit length is
\begin{align}
{\mathfrak{G}_{{\mathrm{c}}}}={\frac{1}{2}}{}{{\mathtt{v}}_{{\mathrm{c}}}}^2=\frac{1}{2}\sigma_c^2.
\end{align}
With the macroscopic energy release rate ${\mathfrak{G}_{rel}}$ defined by
\begin{align}
{\mathfrak{G}_{rel}}={K}^2/(2(1-{\mathcal{V}}^2)^{1/2}),
\end{align}
it is seen that
\begin{align}
{\mathfrak{G}_{rel}}={\mathfrak{G}_{{\mathrm{c}}}}{{\mathfrak{C}}}_{\mathfrak{S}}^{2}.
\end{align}
Thus, the ration of energy spent in bond breaking vs the energy release is ${{\mathfrak{C}}}_{\mathfrak{S}}^{-2}$ (its square root is plotted in Fig. \ref{Fig2}(a) to show the dependence on $\mathcal{V}$).

\section{Mode III crack in  triangular lattice model}
\label{appTaniso}
Consider an infinite triangular lattice consisting of point particles of mass ${M}$. Each particle is connected with six neighbors by the same linearly elastic bonds each of length ${b}$ and spring constant ${K}$, along with the lattice fracture condition. For this lattice model, a mode III crack propagation with subcritical speed is studied. Suppose that a mode III crack is moving at a constant velocity with magnitude ${c_s}>{{\mathtt{V}}}={\frac{1}{2}}{{\mathcal{V}}}{c_s}>0, $ in the crack plane ${\mathcal{P}}$ between ${\mathtt{y}}=0$ and ${\mathtt{y}}=-1.$ The rectangular coordinates described in \cite{Bls4} and \cite{disloc_Bls00} are employed.
Note that ${{\mathcal{V}}}{\mathtt{t}}=2{{\mathtt{V}}}{t}/{b}$ so that the macroscopic moving coordinate ${x}-{{\mathtt{V}}}{t}={\frac{1}{2}}{b}({\mathtt{x}}-{{\mathcal{V}}}{\mathtt{t}})$ as desired. Thus, the crack (scaled) velocity ${\mathcal{V}}=2$ corresponds to the physical velocity ${\mathtt{V}}={c_s}$.
Let
\begin{align}
\mathcal{C}_{\mathfrak{T}}=2.
\label{tgsound}
\end{align}
The displacement field is skew-symmetric following \cite{Bls4} and \cite{Marder}. 
Explicitly, taking into account the piecewise nature of interactions,
\begin{align}
\frac{3}{2}\frac{d^2}{d{\mathtt{t}}^2}{{\mathtt{u}}}_{{\mathtt{x}}, {\mathtt{y}}}{}&=\triangle{{\mathtt{u}}}_{{\mathtt{x}}, {\mathtt{y}}}{}+
{{{\mathscr{H}}}({{\mathcal{V}}}{{\mathtt{t}}}+{\frac{1}{2}}-{\mathtt{x}}){{\updelta}}_{{\mathtt{y}}, 0}}\llbracket {{\mathtt{u}}}\rrbracket^{{\mathcal{P}}{{\mathrel{\text{\ooalign{$\swarrow$\cr$\nearrow$}}}}}}_{\mathtt{x}}({\mathtt{t}})\notag\\
&-{{{\mathscr{H}}}({{\mathcal{V}}}{{\mathtt{t}}}-{\frac{1}{2}}-{\mathtt{x}}+1){{\updelta}}_{{\mathtt{y}}, -1}}\llbracket {{\mathtt{u}}}\rrbracket^{{\mathcal{P}}{{\mathrel{\text{\ooalign{$\searrow$\cr$\nwarrow$}}}}}}_{\mathtt{x}}({\mathtt{t}})\notag\\
&+{{{\mathscr{H}}}({{\mathcal{V}}}{{\mathtt{t}}}-{\frac{1}{2}}-{\mathtt{x}}){{\updelta}}_{{\mathtt{y}}, 0}}\llbracket {{\mathtt{u}}}\rrbracket^{{\mathcal{P}}{{\mathrel{\text{\ooalign{$\searrow$\cr$\nwarrow$}}}}}}_{\mathtt{x}}({\mathtt{t}})\notag\\
&-{{{\mathscr{H}}}({{\mathcal{V}}}{{\mathtt{t}}}+{\frac{1}{2}}-{\mathtt{x}}-1){{\updelta}}_{{\mathtt{y}}, -1}}\llbracket {{\mathtt{u}}}\rrbracket^{{\mathcal{P}}{{\mathrel{\text{\ooalign{$\swarrow$\cr$\nearrow$}}}}}}_{\mathtt{x}}({\mathtt{t}}),
\label{tg_modeIIIcrackeq1}
\end{align}
$\forall ({\mathtt{x}}, {\mathtt{y}})\in{{\mathbb{Z}^2}}.$ 
Assuming ${{\mathtt{u}}}_{\mathtt{x}, \mathtt{y}}(t)={{\mathtt{u}}}_{\mathtt{y}}({\mathtt{z}})$ equation \eqref{tg_modeIIIcrackeq1} can be rewritten in the form 
\begin{align}
\frac{3}{2}{\mathcal{V}}^2\frac{d^2}{d{\mathtt{z}}^2}{{\mathtt{u}}}_{\mathtt{y}}({\mathtt{z}})&={{\mathtt{u}}}_{\mathtt{y}}({\mathtt{z}}+2)+{{\mathtt{u}}}_{\mathtt{y}}({\mathtt{z}}-2)+{{\mathtt{u}}}_{\mathtt{y}+1}({\mathtt{z}}+1)\notag\\
&+{{\mathtt{u}}}_{\mathtt{y}-1}({\mathtt{z}}+1)+{{\mathtt{u}}}_{\mathtt{y}+1}({\mathtt{z}}-1)\notag\\
&+{{\mathtt{u}}}_{\mathtt{y}-1}({\mathtt{z}}-1)-6{{\mathtt{u}}}_{\mathtt{y}}({\mathtt{z}}), 
\label{tg_modeIIIcrackeq2}
\end{align}
away from the crack. 
With $\mathtt{y}=0$,
\begin{align}
\frac{3}{2}{\mathcal{V}}^2\frac{d^2}{d{\mathtt{z}}^2}{{\mathtt{u}}}_{0}({\mathtt{z}})
&={{\mathtt{u}}}_{0}({\mathtt{z}}+2)+{{\mathtt{u}}}_{0}({\mathtt{z}}-2)+{{\mathtt{u}}}_{1}({\mathtt{z}}+1)\notag\\
&-{{\mathtt{u}}}_{0}({\mathtt{z}}+1)+{{\mathtt{u}}}_{1}({\mathtt{z}}-1)-{{\mathtt{u}}}_{0}({\mathtt{z}}-1)-6{{\mathtt{u}}}_{0}({\mathtt{z}})\notag\\
&+ {{\mathscr{H}}}({\frac{1}{2}}-{\mathtt{z}})({{\mathtt{u}}}_{0}({\mathtt{z}})+{{\mathtt{u}}}_{0}({\mathtt{z}}-1))\notag\\
&+{{\mathscr{H}}}(-{\frac{1}{2}}-{\mathtt{z}})({{\mathtt{u}}}_{0}({\mathtt{z}})+{{\mathtt{u}}}_{0}({\mathtt{z}}+1)). 
\end{align}
At the crack surfaces ${\mathtt{z}} < 0$, an external loading is expected to act on the particles $\mathtt{y}=0$ and $-1$, respectively. Let
\begin{subequations}
\begin{align}
{\mathtt{v}}({\mathtt{z}})&{:=}{\mathtt{v}}^{{{\mathrel{\text{\ooalign{$\swarrow$\cr$\nearrow$}}}}}}({\mathtt{z}})=\llbracket {{\mathtt{u}}}\rrbracket^{{\mathcal{P}}{{\mathrel{\text{\ooalign{$\swarrow$\cr$\nearrow$}}}}}}({\mathtt{z}})\\
&={{\mathtt{u}}}_0({\mathtt{z}}+{\frac{1}{2}})-{{\mathtt{u}}}_{-1}({\mathtt{z}}-{\frac{1}{2}})\\
&={{\mathtt{u}}}_{0}({\mathtt{z}}+{\frac{1}{2}})+{{\mathtt{u}}}_{0}({\mathtt{z}}-{\frac{1}{2}}).
\label{tg_vxf}
\end{align}
\end{subequations}
Indeed,
${\mathtt{v}}^{{{\mathrel{\text{\ooalign{$\searrow$\cr$\nwarrow$}}}}}}({\mathtt{z}})=\llbracket {{\mathtt{u}}}\rrbracket^{{\mathcal{P}}{{\mathrel{\text{\ooalign{$\searrow$\cr$\nwarrow$}}}}}}({\mathtt{z}})={{\mathtt{u}}}_{0}({\mathtt{z}}+{\frac{1}{2}})-{{\mathtt{u}}}{-1}({\mathtt{z}}+\frac{3}{2})
={\mathtt{v}}({\mathtt{z}}+1).$ 
Hence
\begin{align}
{\mathtt{v}}^F=\int_{-\infty}^{+\infty}{\mathtt{v}}_{\mathtt{z}}e^{i{\upxi}{\mathtt{z}}}d{\mathtt{z}}
=2\cos{\frac{1}{2}}{\upxi}{\mathtt{u}}_0^F,
\label{tg_vF}
\end{align}
and $\int_{-\infty}^{+\infty}{\mathtt{v}}_{{\mathtt{z}}+1}e^{i{\upxi}{\mathtt{z}}}d{\mathtt{z}}
=e^{-i{\upxi}}2\cos{\frac{1}{2}}{\upxi}{{\mathtt{u}}}_{0}^F.$ 
In terms of the bond lengths ${\mathtt{v}}$, accounting for the far-field stress, 
and using \eqref{tg_vF}, 
\begin{subequations}\begin{eqnarray}
{\mathtt{v}}_{+}+{{\mathpzc{L}}}_{\mathfrak{T}}{\mathtt{v}}_{-}=\frac{1}{2}\frac{1-{{\mathpzc{L}}}_{\mathfrak{T}}}{1+\cos{\upxi}}{{\mathfrak{q}}},\label{tg_WHeq1}\\
\text{where }
{{\mathpzc{L}}}_{\mathfrak{T}}=1-\frac{\cos^2{\frac{1}{2}}{\upxi}}{\cos{\upxi}}(1-{{\mathpzc{L}}}_{0}({\upxi})), {{\mathpzc{L}}}_{0}=\frac{{\mathpzc{h}}}{{\mathpzc{r}}}.\label{tg_Lz}
\end{eqnarray}\label{tg_WGL}\end{subequations}
Recall that ${{{\mathfrak{q}}}}({\upxi})=\frac{{\upsigma}}{i{\upxi}+{0+}}$.
Above is same as the limit of the problem of brittle fracture when number of rows in a finite strip of triangular lattice tends to infinity \cite{disloc_Bls00}; in particular, $${{\mathpzc{L}}}_{\mathfrak{T}}({\upxi})=\frac{\cos{\upxi}-\lambda({\upxi})}{(1+\lambda({\upxi}))\cos{\upxi}}.$$
Also above equation is same as \eqref{hx_WHeqK} except for a different kernel \eqref{hx_LK}. 

It can be shown that 
index ${{\mathpzc{L}}}_{\mathfrak{T}}=0$, ${{\mathpzc{L}}}_{\mathfrak{T}}(\pm\infty)=1,$
and
$${{\mathpzc{L}}}_{\mathfrak{T}}({\upxi})\sim \frac{\sqrt{3}}{4}\sqrt{\mathcal{C}_{\mathfrak{T}}^2-{\mathcal{V}}^2}\sqrt{0+i{\upxi}}\sqrt{0-i{\upxi}}$$ as ${\upxi}\to0$.
The decomposition
\begin{align}
{{\mathpzc{L}}}_{\mathfrak{T}}={{\mathpzc{L}}}_{{\mathfrak{T}}+}{{\mathpzc{L}}}_{{\mathfrak{T}}-}
\end{align}
with
\begin{align}
{{\mathpzc{L}}_{\mathfrak{T}}}_\pm({\upxi})&=\exp(\mp\frac{1}{2\pi i} \int_{\mathbb{R}}\frac{\log {{\mathpzc{L}}_{\mathfrak{T}}}({s})}{{\upxi}-{s}}d{s}), {\upxi}\in{\mathbb{C}}, \Im{\upxi}\gtrless\mp0,\label{tg_LKpmexp}
\end{align}
holds that gives the following asymptotic expressions (${\mathcal{V}}<\mathcal{C}_{\mathfrak{T}}$): 
\begin{subequations}
\begin{align}
{{\mathpzc{L}}}_{{\mathfrak{T}}+}({\upxi}) &\to 1~({\upxi} \to+i\infty),\quad {{\mathpzc{L}}}_{{\mathfrak{T}}-}({\upxi}) \to 1~({\upxi} \to-i\infty), \\
{{\mathpzc{L}}}_{{\mathfrak{T}}+}({\upxi}) &\to (\frac{\sqrt{3}}{4}\sqrt{\mathcal{C}_{\mathfrak{T}}^2-{\mathcal{V}}^2})^{1/2}(0-i\upxi){{\mathfrak{C}}}^{-1}_{\mathfrak{T}}~({\upxi} \to 0),\\{{\mathpzc{L}}}_{{\mathfrak{T}}-}({\upxi}) &\to (\frac{\sqrt{3}}{4}\sqrt{\mathcal{C}_{\mathfrak{T}}^2-{\mathcal{V}}^2})^{1/2}(0+i\upxi){{\mathfrak{C}}}_{\mathfrak{T}}~({\upxi} \to 0), 
\label{tg_Lzplim}
\end{align}
\end{subequations}
where
\begin{align}{{\mathfrak{C}}}_{\mathfrak{T}}&=\exp(\frac{1}{\pi}\int_0^{+\infty}\frac{\arg{{{\mathpzc{L}}}_{\mathfrak{T}}({\upxi})}}{{\upxi}}d{\upxi}).
\label{tg_ConL}
\end{align}
According to \cite{disloc_Bls00}, it is ${{\mathfrak{C}}}_{\mathfrak{T}}^2-1$ which measures the energy radiated per unit broken bond energy while ${{\mathfrak{C}}}_{\mathfrak{T}}^2$ is the corresponding measure of energy release rate. Thus, $1-{{\mathfrak{C}}}_{\mathfrak{T}}^{-2}$ measures the energy radiated per unit energy release while ${{\mathfrak{C}}}_{\mathfrak{T}}^{-2}$ measures the energy spent in fracture per unit energy release (its square root is plotted in Fig. \ref{Fig2}(b) to show the dependence on $\mathcal{V}$). 

A slight generalization is also considered with horizontal bond of spring constant ${\upchi}$. 
Taking into account the anisotropy parameter ${\upchi}$, the right side of \eqref{tg_modeIIIcrackeq2} becomes ${\upchi}{{\mathtt{u}}}_{\mathtt{y}}({\mathtt{z}}+2)+{\upchi}{{\mathtt{u}}}_{\mathtt{y}}({\mathtt{z}}-2)+{{\mathtt{u}}}_{\mathtt{y}+1}({\mathtt{z}}+1)+{{\mathtt{u}}}_{\mathtt{y}-1}({\mathtt{z}}+1)+{{\mathtt{u}}}_{\mathtt{y}+1}({\mathtt{z}}-1)+{{\mathtt{u}}}_{\mathtt{y}-1}({\mathtt{z}}-1)-(4+2{\upchi}){{\mathtt{u}}}_{\mathtt{y}}({\mathtt{z}})$ and
\begin{align}
{{\mathpzc{h}^2}}({\upxi})=\frac{4+2{\upchi}-2{\upchi}\cos2{\upxi}+\frac{3}{2}(0+i{\upxi}{\mathcal{V}})^2}{2\cos {\upxi}}-2.
\label{tg_h2chi}
\end{align}
The crack (scaled) velocity ${\mathcal{V}}=\frac{2}{\sqrt{3}}\sqrt{1+2{{\upchi}}}$ corresponds to the physical velocity equal to the speed of `sound', i.e. ${\mathtt{V}}={c_s}$, hence, \eqref{tgsound} becomes $\mathcal{C}_{\mathfrak{T}}=\frac{2}{\sqrt{3}}\sqrt{1+2{{\upchi}}}$.

\section{Mode III crack in hexagonal lattice}
This section follows the notation of \cite{Bls5} (see also \cite{Bls0,Bls4,Bls5,Bls6}) and several definitions and form of equations for the hexagonal lattice model are analogous to those presented in the context of wave scattering. 
The equation of motion for particles on the crack plane ${\mathcal{P}}$ is obtained by taking into account the snapping bonds between $({\mathtt{x}}, 0)$ and $({\mathtt{x}^{\ast}},-1)$. 

Suppose that $\rho$ is the three dimensional mass density (assuming spacing ${b}$ between parallel sub-lattices), ${{{\upmu}}}$ is the linear elastic shear modulus, and ${{c_s}}$ is the macroscopic shear wave speed. Motivated by the three-dimensional context briefly described above, let ${M}={\frac{1}{2}}\rho {b}^3, {K}=\frac{2}{3}{{{\upmu}}} {{b}}, {x}=\frac{\sqrt{3}}{2}{\mathtt{x}}{b}, {y}=\frac{3}{2}({\mathtt{y}}+\frac{1}{3}){b}, {t}={b} {\mathtt{t}}/{c_s}.$
Via a long-wave approximation, the lattice corresponds to a homogeneous body of density $2{M}/{b}^3$ and shear modulus $3{K}/2{b}$ (the lattice is assumed to be of a unit thickness). Accordingly, the shear wave velocity is given by ${{c_s}} 
=\sqrt{{\upmu}/\rho}=\sqrt{3{K}{b}^2/4{M}}.$

Consider a mode III crack moving at a constant velocity with magnitude ${{\mathtt{V}}}=\frac{\sqrt{3}}{2}{{\mathcal{V}}}{c_s}>0, $ in the crack plane ${\mathcal{P}}$ between ${\mathtt{y}}=-{{\frac{1}{2}}}+{{\frac{1}{2}}}$ and ${{\mathtt{y}^{\ast}}}=-{{\frac{1}{2}}}-{{\frac{1}{2}}}.$
A subsonic mode III crack motion is assumed: $0 < {\mathtt{V}} < {{c_s}}.$ 
Note that ${{\mathcal{V}}}{\mathtt{t}}=\frac{2}{\sqrt{3}}({{\mathtt{V}}}/{c_s}){c_s}{t}/{b}=\frac{2}{\sqrt{3}}{{\mathtt{V}}}{t}/{b}$ so that the macroscopic moving coordinate ${x}-{{\mathtt{V}}}{t}=\frac{\sqrt{3}}{2}{b}({\mathtt{x}}-{{\mathcal{V}}}{\mathtt{t}})$ as desired. Thus, the crack (scaled) velocity ${\mathcal{V}}=\frac{2}{\sqrt{3}}$ corresponds to the physical velocity equal to the speed of `sound', i.e. ${\mathtt{V}}={c_s}$. Let
\begin{align}
\mathcal{C}_{\mathfrak{H}}=\frac{2}{\sqrt{3}}.
\label{hxsound}
\end{align}
The equation of motion for the particle located at $({\mathtt{x}}, {\mathtt{y}})$, based on the classical (Newtonian) mechanics, is directly affected by the presence of crack whenever ${\mathtt{y}}=-{{\frac{1}{2}}}\pm{{\frac{1}{2}}}$. 
The lattice fracture criterion is
\begin{align}
\llbracket {{\mathtt{u}}}\rrbracket^{{\mathcal{P}}{\mathrel{\text{\ooalign{$\downarrow$\cr$\uparrow$}}}}}({\mathtt{x}}, {{\mathtt{t}}})={{\mathtt{u}}}_{{\mathtt{x}}, {0}}(t)-{{\mathtt{u}}}_{{\mathtt{x}^{\ast}}, {-1}}(t)={{\mathtt{v}}_{{\mathrm{c}}}}, {{\mathtt{t}}}=\frac{\mathtt{x}}{{\mathcal{V}}}.
\label{hx_fracbond}
\end{align}
Explicitly, taking into account the piecewise nature of interactions and lattice fracture condition, the forces compensating the interaction must be introduced for ${\mathtt{z}} < 0$,
\begin{subequations}\begin{eqnarray}\frac{3}{4}\frac{d^2}{d{\mathtt{t}}^2}{{\mathtt{u}}}_{{\mathtt{x}}, {\mathtt{y}}}{}&=&{{\mathtt{u}^{\ast}}}_{{\mathtt{x}^{\ast}}+1, {\mathtt{y}^{\ast}}}{}+{{\mathtt{u}^{\ast}}}_{{\mathtt{x}^{\ast}}-1, {\mathtt{y}^{\ast}}}{}+{{\mathtt{u}^{\ast}}}_{{\mathtt{x}^{\ast}}, {\mathtt{y}^{\ast}}-1}-3{{\mathtt{u}}}_{{\mathtt{x}}, {\mathtt{y}}}{}\notag\\
&&+{{\mathscr{H}}}({{\mathcal{V}}}{{\mathtt{t}}}-{\mathtt{x}}){{\updelta}}_{{\mathtt{y}}, 0}({{\mathtt{u}}}_{{\mathtt{x}}, {\mathtt{y}}}{}-{{\mathtt{u}^{\ast}}}_{{\mathtt{x}^{\ast}}, {\mathtt{y}^{\ast}}-1}), \\\frac{3}{4}\frac{d^2}{d{\mathtt{t}}^2}{{\mathtt{u}^{\ast}}}_{{\mathtt{x}^{\ast}}, {\mathtt{y}^{\ast}}}{}&=&{{\mathtt{u}}}_{{\mathtt{x}}+1, {\mathtt{y}}}{}+{{\mathtt{u}}}_{{\mathtt{x}}-1, {\mathtt{y}}}{}+{{\mathtt{u}}}_{{\mathtt{x}}, {\mathtt{y}}+1}-3{{\mathtt{u}^{\ast}}}_{{\mathtt{x}^{\ast}}, {\mathtt{y}^{\ast}}}{}\notag\\
&&+{{\mathscr{H}}}({{\mathcal{V}}}{{\mathtt{t}}}-{\mathtt{x}^{\ast}}){{\updelta}}_{{\mathtt{y}^{\ast}},-1}({{\mathtt{u}^{\ast}}}_{{\mathtt{x}^{\ast}}, {\mathtt{y}^{\ast}}}{}-{{\mathtt{u}}}_{{\mathtt{x}}, {\mathtt{y}}+1}),\label{hx_dnewton}\end{eqnarray}\label{hx_dnewtonfull}\end{subequations}for $ ({\mathtt{x}}, {\mathtt{y}})\in{{\mathbb{Z}^2}}$ and $({\mathtt{x}^{\ast}}, {\mathtt{y}^{\ast}})\in{{\mathbb{Z}^2}}$, respectively, 
such that $({\mathtt{x}}, {\mathtt{y}}), ({{\mathtt{x}^{\ast}}}, {{\mathtt{y}^{\ast}}})$ lie away from 
the boundary, the nature of which is discussed below for each case studied in this paper. 

At macroscale, the far field {\em boundary} conditions (`at $\mathtt{y}=\pm\infty$') are such that a homogeneous shear stress ${\upsigma}$ is applied. In order to improve the clarity of expressions, instead of writing $\frac{{\upsigma}}{{\upmu}}$, simply ${\upsigma}$ is used.

For the considered steady-state problem a moving coordinate, ${\mathtt{z}}=\mathtt{x}-{\mathcal{V}}{\mathtt{t}}$ is introduced. Assuming ${{\mathtt{u}}}_{\mathtt{x}, \mathtt{y}}(t)={{\mathtt{u}}}_{\mathtt{y}}({\mathtt{z}})$ equation \eqref{hx_dnewtonfull} can be rewritten in the form
\begin{subequations}\begin{eqnarray}
\frac{3}{4}{\mathcal{V}}^2\frac{d^2}{d{\mathtt{z}}^2}{{\mathtt{u}}}_{{\mathtt{y}}}({\mathtt{z}})&={{\mathtt{u}^{\ast}}}_{{\mathtt{y}^{\ast}}}({\mathtt{z}}+1)+{{\mathtt{u}^{\ast}}}_{{\mathtt{y}^{\ast}}}({\mathtt{z}}-1)+{{\mathtt{u}^{\ast}}}_{{\mathtt{y}^{\ast}}-1}({\mathtt{z}})-3{{\mathtt{u}}}_{{\mathtt{y}}}({\mathtt{z}})\notag\\
&+{{\mathscr{H}}}(-{\mathtt{z}}){\updelta}_{\mathtt{y}, 0}\bigr({{\mathtt{u}}}_{{\mathtt{y}}}({\mathtt{z}})-{{\mathtt{u}^{\ast}}}_{{\mathtt{y}^{\ast}}-1}({\mathtt{z}})\bigr), \label{hx_modeIIIcrackeq}\\
\frac{3}{4}{\mathcal{V}}^2\frac{d^2}{d{\mathtt{z}}^2}{{\mathtt{u}^{\ast}}}_{{\mathtt{y}^{\ast}}}({\mathtt{z}})&={{\mathtt{u}}}_{{\mathtt{y}}}({\mathtt{z}}+1)+{{\mathtt{u}}}_{{\mathtt{y}}}({\mathtt{z}}-1)+{{\mathtt{u}}}_{{\mathtt{y}}+1}({\mathtt{z}})-3{{\mathtt{u}^{\ast}}}_{{\mathtt{y}^{\ast}}}({\mathtt{z}})\notag\\
&+{{\mathscr{H}}}(-{\mathtt{z}}){\updelta}_{\mathtt{y}^{\ast},-1}\bigr({{\mathtt{u}^{\ast}}}_{{\mathtt{y}^{\ast}}}({\mathtt{z}})-{{\mathtt{u}}}_{{\mathtt{y}}+1}({\mathtt{z}})\bigr).
\label{hx_pmodeIIIcrackeq}
\end{eqnarray}\label{hx_modeIIIcrackeqfull}\end{subequations}
From the Fourier transform ($u^F({\upxi})=\int_{-\infty}^{+\infty}u({\mathtt{z}})e^{i{\upxi}{\mathtt{z}}}d{\mathtt{z}}$),
it immediately follows that
\begin{subequations}
\begin{align}
({{\mathpzc{h}^2}}({\upxi})+2){{\mathtt{u}}}_{\mathtt{y}}^F({\upxi})-({{\mathtt{u}}}_{{\mathtt{y}}+1}^F({\upxi})+{{\mathtt{u}}}_{{\mathtt{y}}-1}^F({\upxi}))=0,
\label{hx_uFeqn}
\end{align}
for all $\mathtt{y}\in\mathbb{Z}\setminus\{0\}$ for mode III crack, where the complex function ${{\mathpzc{h}^2}}$ is defined by
\begin{align}
{{\mathpzc{h}^2}}({\upxi})&{:=}\frac{(2+\frac{3}{4}(0+i{\upxi}{\mathcal{V}})^2)(4+\frac{3}{4}(0+i{\upxi}{\mathcal{V}})^2)}{2\cos{\upxi}}-2\cos{\upxi}-2.
\label{hx_h2}
\end{align}
\label{hx_slepyaneqns}\end{subequations}
The same equation, as described by \eqref{hx_slepyaneqns}, is satisfied by ${{\mathtt{u}^{\ast}}}_{{\mathtt{y}^{\ast}}}^F$ for all ${\mathtt{y}^{\ast}}\in\mathbb{Z}\setminus\{-1\}$. For the problem of a mode III crack in an unbounded lattice, it is assumed that the lattice waves are radiated away from the mode III crack tip.
Hence, the solution of \eqref{hx_uFeqn} is expressed as 
\begin{subequations}
\begin{eqnarray*}
{{\mathtt{u}}}_{\mathtt{y}}^F({\upxi})=\begin{cases}{{\mathtt{u}}}_0^F({\upxi}){{\lambda}}^{\mathtt{y}}({\upxi})~&({\mathtt{y}} \ge 0), \\
{{\mathtt{u}}}^F_{-1}({\upxi}){{\lambda}}^{-({\mathtt{y}}+1)}({\upxi})~&({\mathtt{y}} \le-1),
\end{cases}\label{hx_ubK}\\
\text{where } {\lambda}({\upxi})=\frac{{\mathpzc{r}}({\upxi})-{\mathpzc{h}}({\upxi})}{{\mathpzc{r}}({\upxi})+{\mathpzc{h}}({\upxi})}~(|{\lambda}| \le 1).\label{hx_lam}
\end{eqnarray*}
Similarly, since ${{\mathtt{u}^{\ast}}}_{{\mathtt{y}^{\ast}}}^F$ satisfies the same equation as that for ${{\mathtt{u}}}_{\mathtt{y}}^F,$ 
\begin{align}
{{\mathtt{u}^{\ast}}}_{{\mathtt{y}^{\ast}}}^F({\upxi})=\begin{cases}{{\mathtt{u}^{\ast}}}_0^F({\upxi}){{\lambda}}^{{\mathtt{y}^{\ast}}}({\upxi})~&({\mathtt{y}^{\ast}} \ge 0), \\
{{\mathtt{u}^{\ast}}}^F_{-1}({\upxi}){{\lambda}}^{-({{\mathtt{y}^{\ast}}}+1)}({\upxi})~&({\mathtt{y}^{\ast}} \le-1).
\end{cases}
\label{hx_pubK}
\end{align}
\label{hx_ubulkK}\end{subequations}
The skew-symmetry of the solution follows \cite{Bls5} in the sense that 
\begin{align}
{{\mathtt{u}}}_{\mathtt{y}}({\mathtt{z}})=-{{\mathtt{u}^{\ast}}}_{-\mathtt{y}-1}({\mathtt{z}}), {{\mathtt{u}^{\ast}}}_{\mathtt{y}}({\mathtt{z}})=-{{\mathtt{u}}}_{-\mathtt{y}-1}({\mathtt{z}}), \mathtt{y}\ge0.
\label{hx_skewcond}
\end{align}
holds.
Following \cite{Bls5},
\begin{align}
{{\mathtt{u}^{\ast}}}^F_{0}={{\mathpzc{N}}}{{\lambda}}{{\mathtt{u}}}^F_{0},
\label{hx_up0F}
\text{where }
{{\mathpzc{N}}}({\upxi})=\frac{2\cos{\upxi}{{\lambda}}({\upxi})^{-1}+1}{3(1+\frac{1}{4}(0+i{\upxi}{\mathcal{V}})^2)},
\end{align}
and
\begin{align}
{{\mathtt{u}}}^F_{{-1}}=-{{\mathpzc{N}}}{{\lambda}}{{\mathtt{u}}}^F_{0}.
\label{hx_un1F}
\end{align}
The equation satisfied by the displacement ${\mathtt{u}}$ at $\mathtt{y}=0$ is
\begin{align}
\frac{3}{4}{\mathcal{V}}^2{{\mathtt{u}}}''_{0}({\mathtt{z}})+({{\mathtt{u}}}_{0}({\mathtt{z}})-{{\mathtt{u}^{\ast}}}_{-1}({\mathtt{z}})){\mathscr{H}}({\mathtt{z}})\\
={{\mathtt{u}^{\ast}}}_{0}({\mathtt{z}}+1)+{{\mathtt{u}^{\ast}}}_{0}({\mathtt{z}}-1)-2{{\mathtt{u}}}_{0}({\mathtt{z}}).
\label{hx_y0}
\end{align}
An analoguous equation is satisfied by ${{\mathtt{u}^{\ast}}}$ at $\mathtt{y}^{\ast}=-1$. For convenience, ${\mathtt{u}}$ is written in place of ${{\mathtt{u}}}_{0}$.
After taking Fourier transform ${{\mathtt{u}}}_{\pm}=\int_{0(-\infty)}^{\infty(0)} {\mathtt{u}}({\mathtt{z}})e^{i{\upxi}{\mathtt{z}}}d{\mathtt{z}},$ of the equation 
for $\mathtt{y}=0$ it is found that
\begin{subequations}\begin{eqnarray}
{{\mathtt{u}}}_{+}({\upxi})+{{\mathpzc{L}}_{\mathfrak{H}}}({\upxi}){{\mathtt{u}}}_{-}({\upxi})
&=&
{\frac{1}{2}}(1-{{\mathpzc{L}}}_{\mathfrak{H}}({\upxi})){\mathfrak{q}}, 
\label{hx_WHeqK}\\
\text{where }
{{\mathpzc{L}}}_{\mathfrak{H}}({\upxi})&=&\frac{{{\mathpzc{N}}}({\upxi})-1}{{{\mathpzc{N}}}({\upxi})+1},\label{hx_LK}
\end{eqnarray}\label{hx_WHeqKfull}\end{subequations}
for ${\upxi}\in\mathbb{R}$ and ${\mathpzc{N}}$ given by \eqref{hx_up0F}. 
Recall that ${{{\mathfrak{q}}}}({\upxi})=\frac{{\upsigma}}{i{\upxi}+{0+}}$.

\begin{figure}[!htb]
\begin{center}
{\includegraphics[width=.6\linewidth]{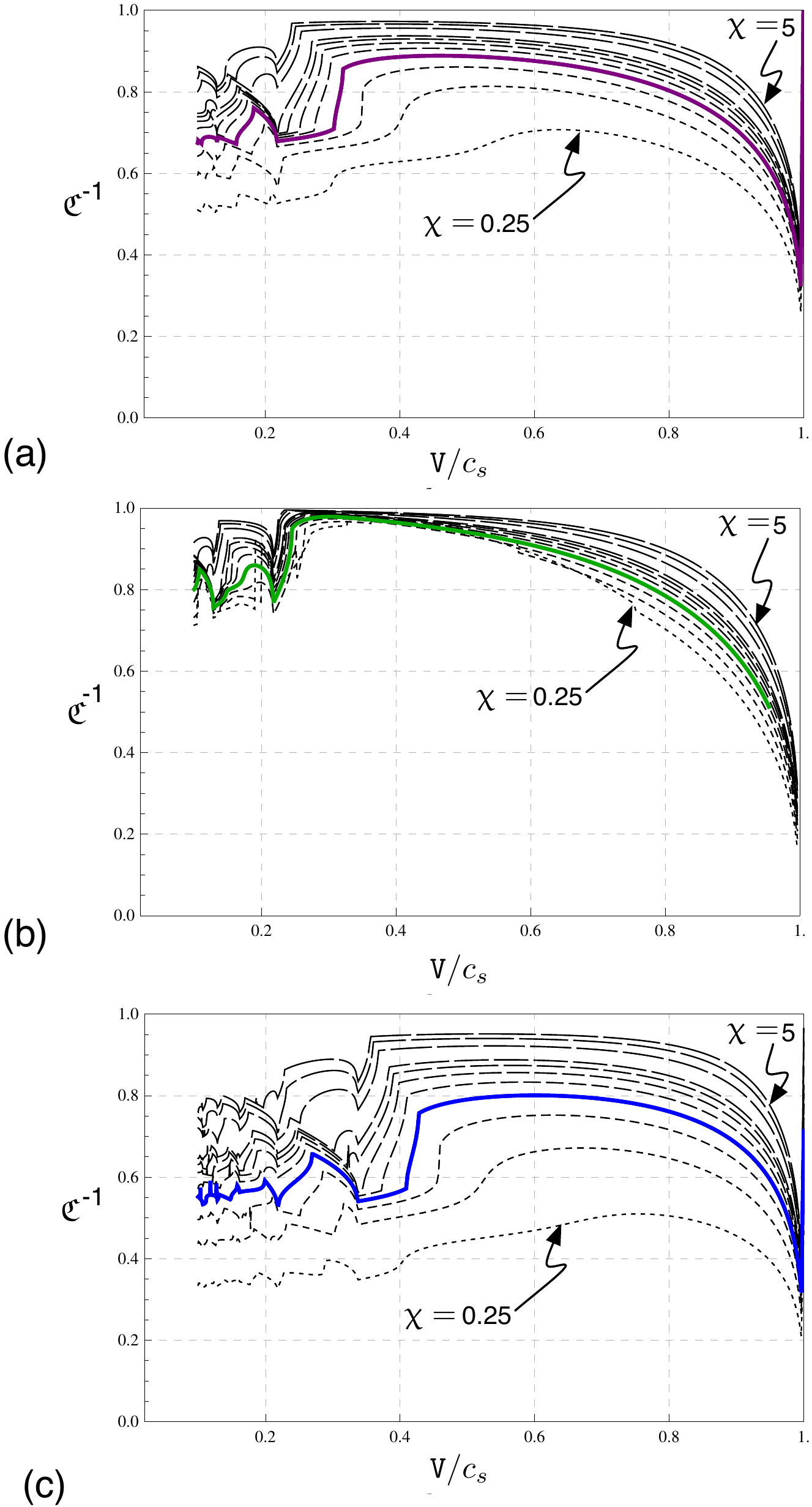}}
\caption{The relation ${{\mathfrak{C}}}^{-1}$ vs ${{\mathtt{V}}}/{c_s}$ for a mode III crack in square (a), triangular (b), and hexagonal (c) lattices where ${\upchi}$ takes values in $S_{{\upchi}}$ \eqref{Schi}. For square (black), triangular (gray), and hexagonal (blue) lattices ${\upchi}=1$ is indicated separately.}
\label{Fig2}
\end{center}
\end{figure}

\begin{figure}[!htb]
\begin{center}
{(a)}{\includegraphics[width=.6\linewidth]{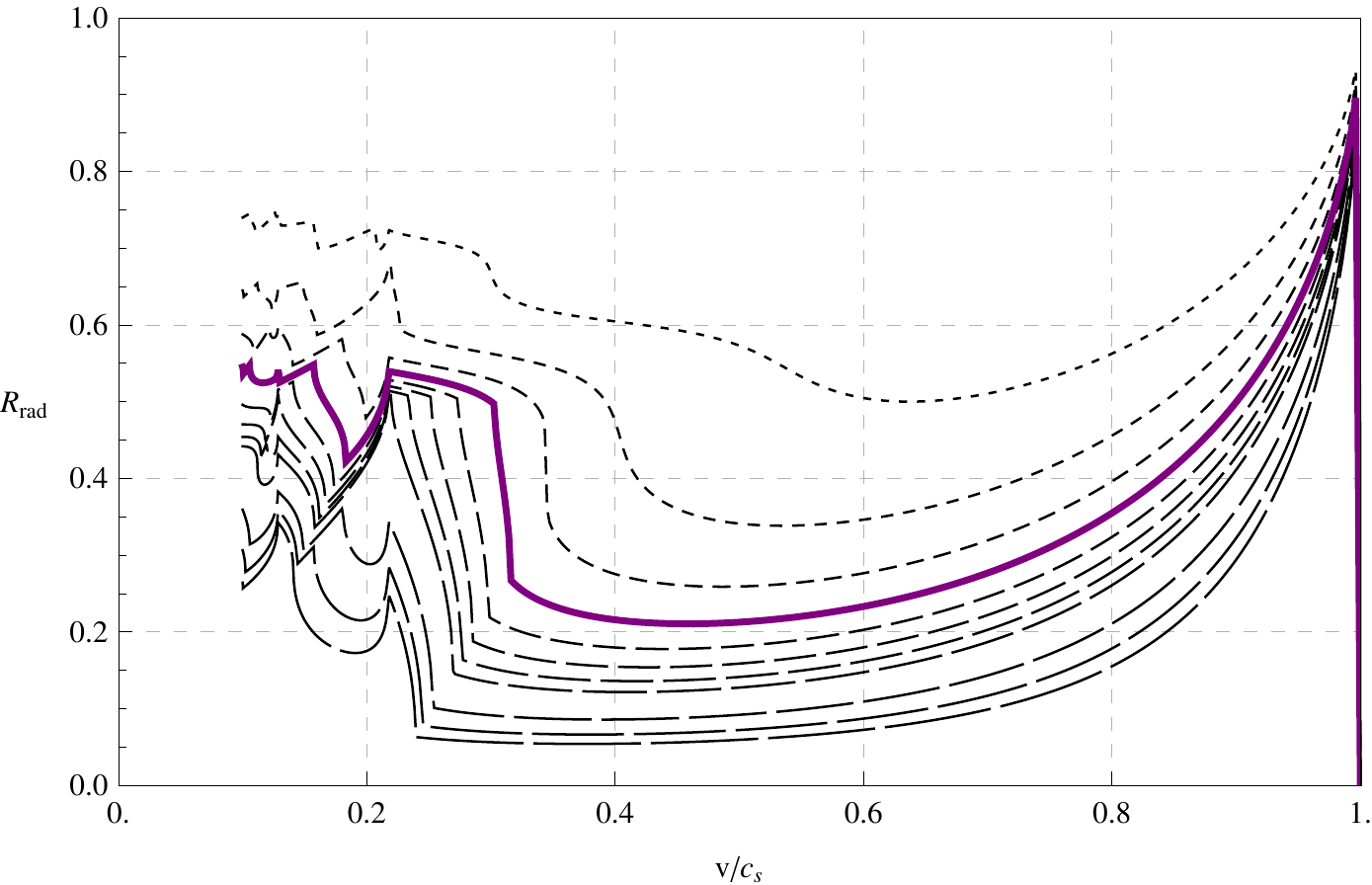}}\\
{(b)}{\includegraphics[width=.6\linewidth]{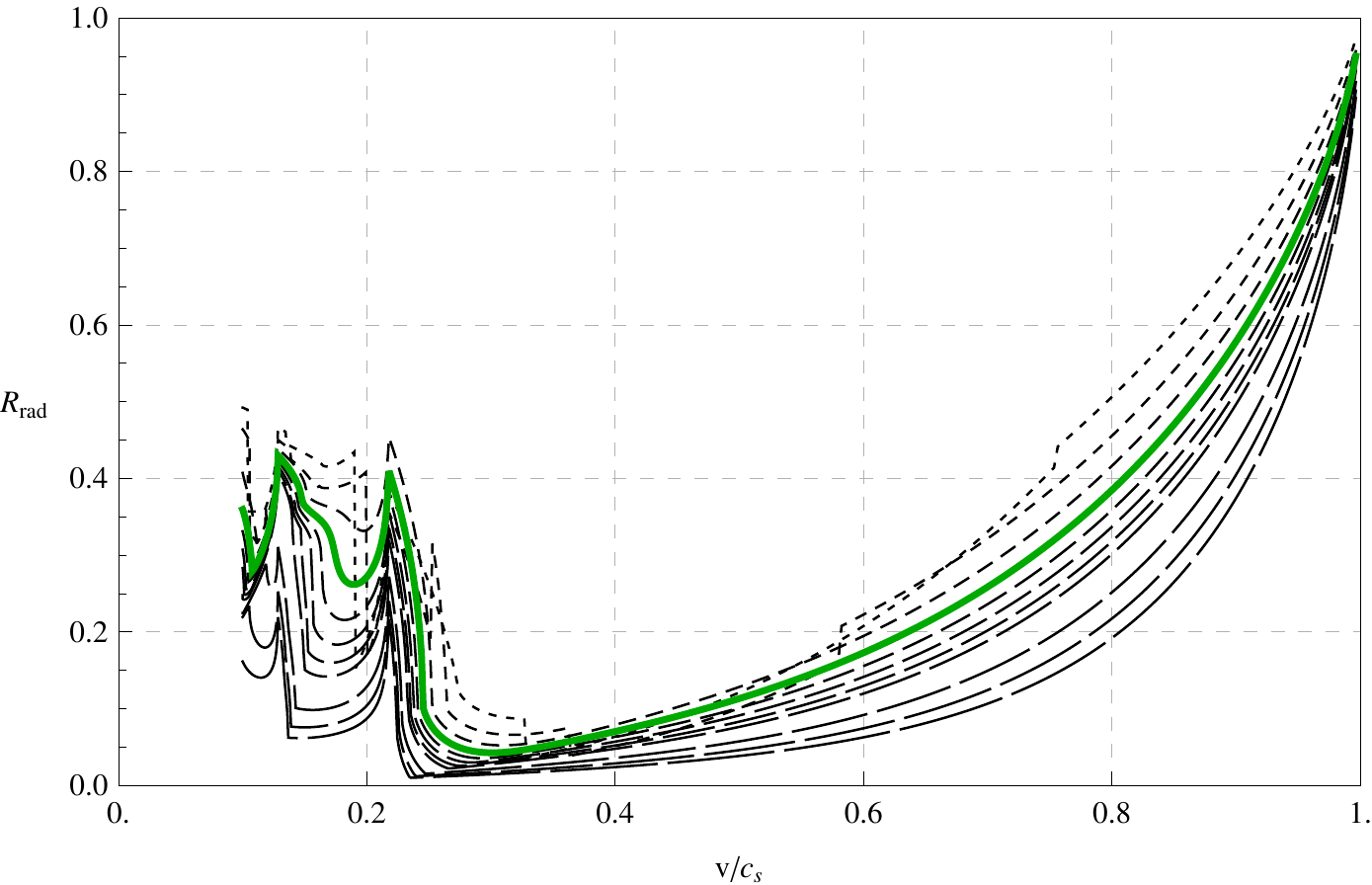}}\\
{(c)}{\includegraphics[width=.6\linewidth]{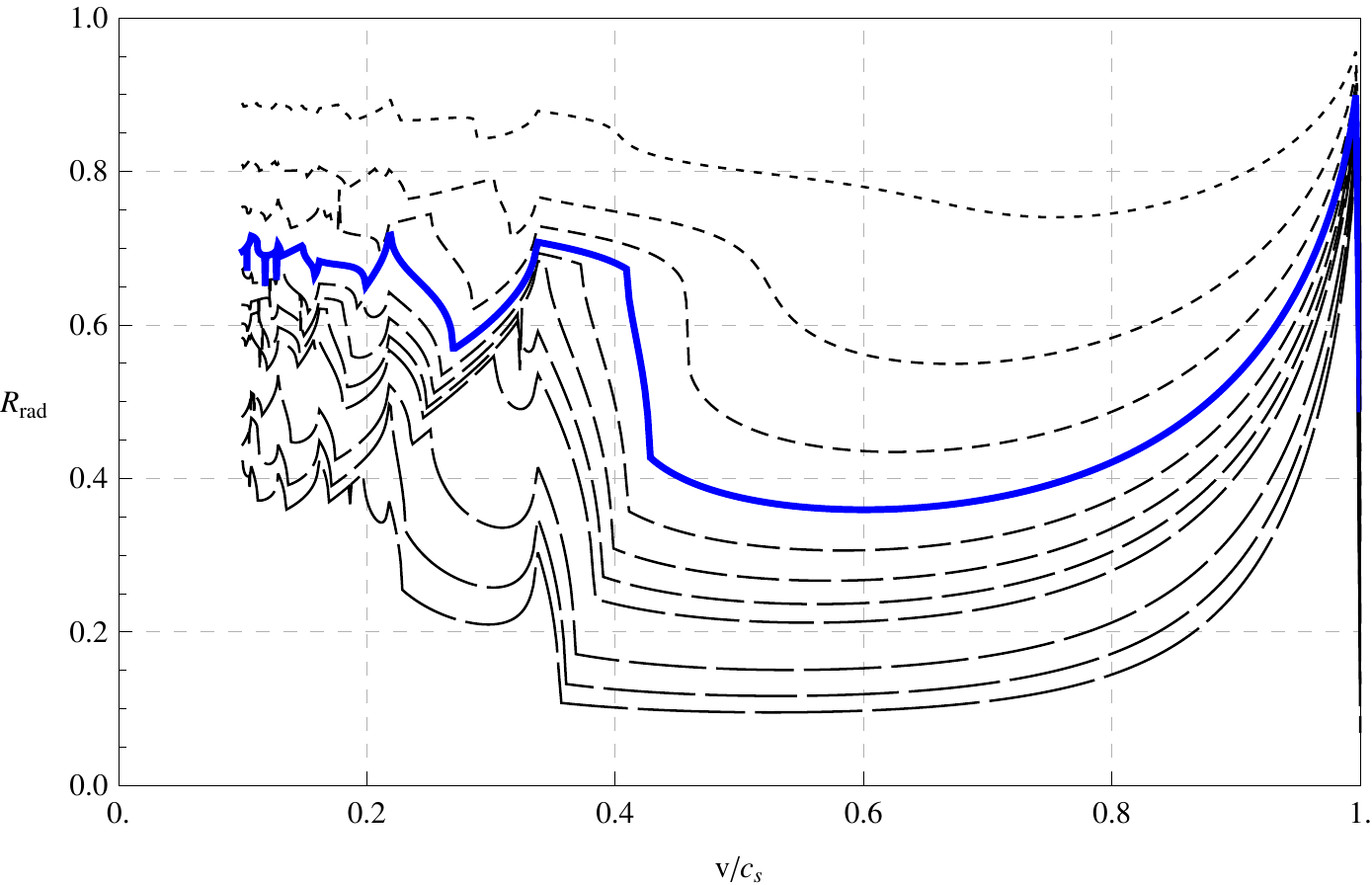}}
\caption{The relation $1-{\mathfrak{G}_{{\mathrm{c}}}}/{\mathfrak{G}_{rel}}$ ($=1-{{\mathfrak{C}}}_{\mathfrak{S}}^{-2}$) vs ${{\mathtt{V}}}/{c_s}$ for a mode III crack in square (a), triangular (b), and hexagonal (c) lattices where ${\upchi}$ takes values in $S_{{\upchi}}$ \eqref{Schi}. For square (black), triangular (gray), and hexagonal (blue) lattices ${\upchi}=1$ is indicated separately.}
\label{Fig3}
\end{center}
\end{figure}

A slight generalization by taking into account an anisotropy parameter ${\upchi}$, the right sides of \eqref{hx_modeIIIcrackeqfull} become ${\upchi}{{\mathtt{u}^{\ast}}}_{{\mathtt{y}^{\ast}}}({\mathtt{z}}+1)+{\upchi}{{\mathtt{u}^{\ast}}}_{{\mathtt{y}^{\ast}}}({\mathtt{z}}-1)+{{\mathtt{u}^{\ast}}}_{{\mathtt{y}^{\ast}}-1}({\mathtt{z}})-(1+2{\upchi}){{\mathtt{u}}}_{{\mathtt{y}}}({\mathtt{z}})$, etc, and in place of \eqref{hx_h2}
\begin{align}
{{\mathpzc{h}^2}}({\upxi})=\frac{(2{\upchi}+\frac{3}{4}(0+i{\upxi}{\mathcal{V}})^2)(2+2{\upchi}+\frac{3}{4}(0+i{\upxi}{\mathcal{V}})^2)}{2{\upchi}\cos{\upxi}}-2{\upchi}\cos{\upxi}-2.
\label{hx_h2chi}
\end{align}
The crack (scaled) velocity ${\mathcal{V}}=\frac{2}{\sqrt{3}}\sqrt{{{\upchi}}}$ corresponds to the physical velocity equal to the speed of `sound', i.e. ${\mathtt{V}}={c_s}$ so that \eqref{hxsound} becomes
$\mathcal{C}_{\mathfrak{H}}=\frac{2}{\sqrt{3}}\sqrt{{{\upchi}}}.$
The definition of the factor ${{\mathpzc{N}}}$ in \eqref{hx_up0F} becomes
$${{\mathpzc{N}}}({\upxi})=({2{{\upchi}}\cos{\upxi}{{\lambda}}({\upxi})^{-1}+1})/({1+2{{\upchi}}+\frac{3}{4}(0+i{\upxi}{\mathcal{V}})^2}).$$
It can be shown that 
index ${{\mathpzc{L}}}_{\mathfrak{H}}=0$, ${{\mathpzc{L}}}_{\mathfrak{H}}(\pm\infty)=1,$
and
$${{\mathpzc{L}}}_{\mathfrak{H}}({\upxi})\sim \frac{1}{2}\sqrt{\mathcal{C}_{\mathfrak{H}}^2-{\mathcal{V}}^2}\sqrt{0+i{\upxi}}\sqrt{0-i{\upxi}}$$ as ${\upxi}\to0$ (recall \eqref{hxsound}).
The decomposition
\begin{align}
{{\mathpzc{L}}}_{\mathfrak{H}}={{\mathpzc{L}}}_{{\mathfrak{H}}+}{{\mathpzc{L}}}_{{\mathfrak{H}}-}
\end{align}
with
\begin{align}
{{\mathpzc{L}}_{\mathfrak{H}}}_\pm({\upxi})&=\exp(\mp\frac{1}{2\pi i} \int_{\mathbb{R}}\frac{\log {{\mathpzc{L}}_{\mathfrak{H}}}({s})}{{\upxi}-{s}}d{s}), {\upxi}\in{\mathbb{C}}, \Im{\upxi}\gtrless\mp0,\label{hx_LKpmexp}
\end{align}
holds that gives the following asymptotic expressions (${\mathcal{V}}<\mathcal{C}_{\mathfrak{H}}$): 
\begin{subequations}
\begin{align}
{{\mathpzc{L}}}_{{\mathfrak{H}}+}({\upxi}) &\to 1~({\upxi} \to+i\infty),\quad {{\mathpzc{L}}}_{{\mathfrak{H}}-}({\upxi}) \to 1~({\upxi} \to-i\infty), \\
{{\mathpzc{L}}}_{{\mathfrak{H}}+}({\upxi}) &\to (\frac{1}{2}\sqrt{\mathcal{C}_{\mathfrak{H}}^2-{\mathcal{V}}^2})^{1/2}(0-i\upxi){{\mathfrak{C}}}_{\mathfrak{H}}^{-1}~({\upxi} \to 0),\\{{\mathpzc{L}}}_{{\mathfrak{H}}-}({\upxi}) &\to (\frac{1}{2}\sqrt{\mathcal{C}_{\mathfrak{H}}^2-{\mathcal{V}}^2})^{1/2}(0+i\upxi){{\mathfrak{C}}}_{\mathfrak{H}}~({\upxi} \to 0),
\label{hx_Lzplim}
\end{align}
\end{subequations}
where
\begin{align}
{{\mathfrak{C}}}_{\mathfrak{H}}&=\exp(\frac{1}{\pi}\int_0^{+\infty}\frac{\arg{{{\mathpzc{L}}}_{\mathfrak{H}}({\upxi})}}{{\upxi}}d{\upxi}).
\label{hx_ConL}
\end{align}
Rest of the analysis is same as that for the square lattice model.

It is ${{\mathfrak{C}}}_{\mathfrak{H}}^2-1$ which measures the energy radiated per unit broken bond energy while ${{\mathfrak{C}}}_{\mathfrak{H}}^2$ is the corresponding measure of energy release rate. Thus, $1-{{\mathfrak{C}}}_{\mathfrak{H}}^{-2}$ measures the energy radiated per unit energy release while ${{\mathfrak{C}}}_{\mathfrak{H}}^{-2}$ measures the energy spent in fracture per unit energy release (its square root is plotted in Fig. \ref{Fig2}(c) to show the dependence on $\mathcal{V}$).

\section{Comparison between square, triangular, and hexagonal Lattices}
A graphical comparison of ${{\mathfrak{C}}}^{-1}$ vs ${{\mathtt{V}}}/{c_s}$ is presented between three types of lattices: square \cite{Slepyan1981a, Slepyanbook}, triangular (as a limit of number of rows tending to infinity from \cite{Marder}), and hexagonal.
For various values of ${{\upchi}}$ in the set
\begin{align}
S_{{\upchi}}=\{0.25, 0.5, 0.75, 1, 1.25, 1.5, 1.75, 2, 3, 4, 5 \},
\label{Schi}
\end{align}
the plots of ${{\mathfrak{C}}}^{-1}$ vs ${{\mathtt{V}}}/{c_s}$ are presented in Fig. \ref{Fig2}(a), (b), and (c) for square, triangular, and hexagonal lattice, respectively.

Fig. \ref{Fig3} presents $1-{{\mathfrak{C}}}^{-2}$ vs ${{\mathtt{V}}}/{c_s}$ in parts (a), (b), and (c) for square, triangular, and hexagonal lattice, respectively.

\section*{Acknowledgments}
The support of SERB MATRICS grant MTR/2017/000013 is gratefully acknowledged.

\renewcommand*{\bibfont}{\footnotesize}
\printbibliography
\end{document}